\documentclass[twocolumn,superscriptaddress,pra,longbibliography,preprintnumbers,amsmath,amssymb,floatfix]{revtex4-1}
\usepackage{upgreek}
\pdfoutput=1

\usepackage{amsmath, amsthm, amssymb, amsfonts}

\usepackage{siunitx}
\usepackage{lmodern}

\usepackage[utf8]{inputenc}	
\usepackage[T1]{fontenc}	
\usepackage{xcolor}		
\usepackage[colorlinks = true,
linkcolor = purple,
urlcolor  = blue,
citecolor = cyan,
anchorcolor = black]{hyperref}	
\usepackage{microtype}		
\usepackage{float}			
\usepackage{graphicx}

\newcommand{\figref}[2]{Fig.~\ref{#1}{(#2)}}
\newcommand{\figrefs}[1]{Fig.~\ref{#1}}

\begin{document}
\title{Femtosecond streaking in ambient air}

\author{A.~Korobenko}
\affiliation{Joint Attosecond Science Laboratory, National Research Council of Canada and University of Ottawa, Ottawa, Ontario K1A0R6, Canada}
\author{K.~Johnston}
\affiliation{Joint Attosecond Science Laboratory, National Research Council of Canada and University of Ottawa, Ottawa, Ontario K1A0R6, Canada}
\author{M.~Kubullek}
\affiliation{Physics Department, Ludwig-Maximilians-Universität Munich, Am Coulombwall 1, 85748 Garching, Germany}

\author{L.~Arissian}
\affiliation{Joint Attosecond Science Laboratory, National Research Council of Canada and University of Ottawa, Ottawa, Ontario K1A0R6, Canada}
\affiliation{National Research Council Canada, 100 Sussex Dr., Ottawa, Ontario K1A 0R6, Canada}
\author{Z.~Dube}
\affiliation{Joint Attosecond Science Laboratory, National Research Council of Canada and University of Ottawa, Ottawa, Ontario K1A0R6, Canada}
\author{T.~Wang}
\affiliation{Joint Attosecond Science Laboratory, National Research Council of Canada and University of Ottawa, Ottawa, Ontario K1A0R6, Canada}
\author{M.~K\"ubel}
\affiliation{Joint Attosecond Science Laboratory, National Research Council of Canada and University of Ottawa, Ottawa, Ontario K1A0R6, Canada}
\affiliation{Institute for Optics and Quantum Electronics, University of Jena, Max-Wien-Platz 1, 07743 Jena, Germany}
\author{A.Yu.~Naumov}
\affiliation{Joint Attosecond Science Laboratory, National Research Council of Canada and University of Ottawa, Ottawa, Ontario K1A0R6, Canada}
\author{D.M.~Villeneuve}
\affiliation{Joint Attosecond Science Laboratory, National Research Council of Canada and University of Ottawa, Ottawa, Ontario K1A0R6, Canada}
\author{M.F.~Kling}
\affiliation{Physics Department, Ludwig-Maximilians-Universität Munich, Am Coulombwall 1, 85748 Garching, Germany}
\affiliation{Max-Planck-Institut für Quantenoptik, Hans-Kopfermann-Straße 1, 85748 Garching, Germany}

\author{P.B.~Corkum}
\affiliation{Joint Attosecond Science Laboratory, National Research Council of Canada and University of Ottawa, Ottawa, Ontario K1A0R6, Canada}
\author{A.~Staudte}
\affiliation{Joint Attosecond Science Laboratory, National Research Council of Canada and University of Ottawa, Ottawa, Ontario K1A0R6, Canada}
\author{B.~Bergues}
\email[Corresponding author: ]{boris.bergues@mpq.mpg.de}
\affiliation{Joint Attosecond Science Laboratory, National Research Council of Canada and University of Ottawa, Ottawa, Ontario K1A0R6, Canada}
\affiliation{Physics Department, Ludwig-Maximilians-Universität Munich, Am Coulombwall 1, 85748 Garching, Germany}
\affiliation{Max-Planck-Institut für Quantenoptik, Hans-Kopfermann-Straße 1, 85748 Garching, Germany}





\begin{abstract}

We demonstrate a novel method to measure the temporal evolution of electric fields with optical frequencies. Our technique is based 
on the detection of transient currents in air plasma. These directional currents result from sub-cycle ionization of 
air with a short pump pulse, and the steering of the released electrons with the pulse to be sampled. We assess the validity 
of our approach by comparing it with different state-of-the-art laser-pulse characterization techniques. Notably, our method works in 
ambient air and facilitates a direct measurement of the field waveform, which can be viewed in real time on an oscilloscope in the 
exact same way as a radio frequency signal.

\end{abstract}

\maketitle
\section{Introduction}

\begin{figure}[ht!]
	\centering\includegraphics{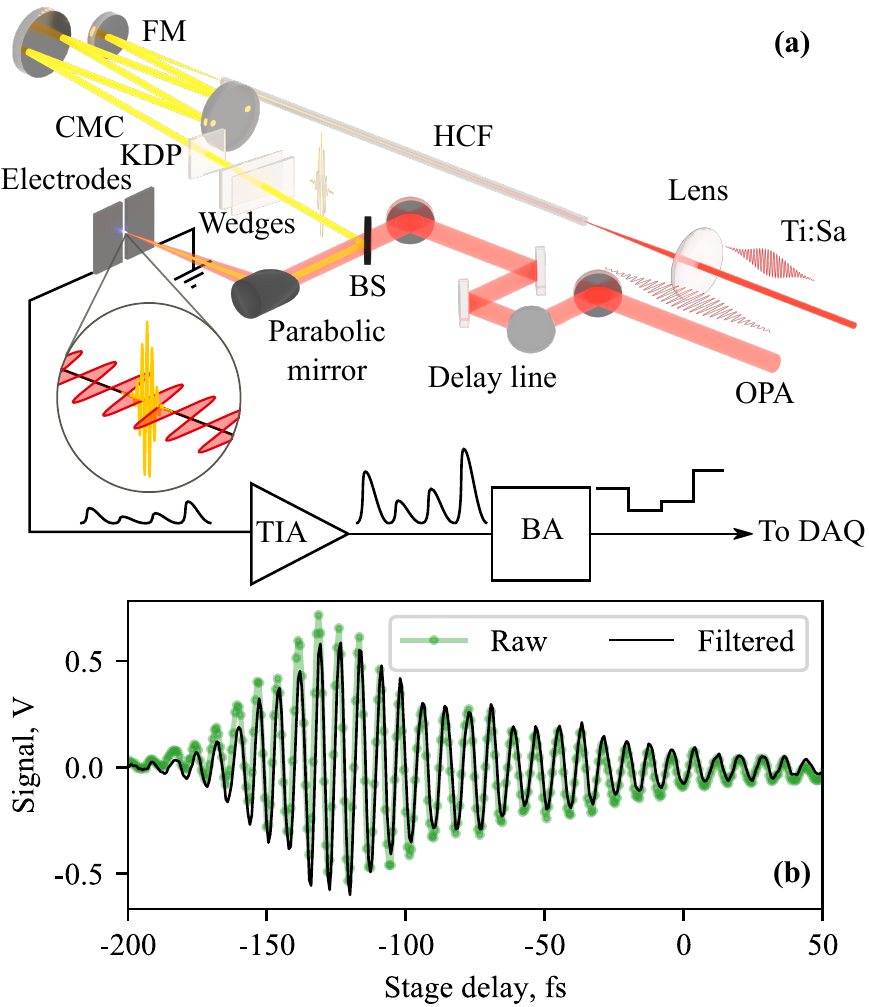}
	\caption{{\bf (a)} Experimental setup. An output of a Ti:Sapphire laser (Ti:Sa) is passed through an Ar-filled hollow-core fiber (HCF), recollimated with a silver focussing mirror  (FM) and compressed in a chirped-mirror compressor (CMC). It is combined with an idler of an OPA on a silicon beam splitter (BS), and focussed by an off-axis parabolic mirror in between a pair of metal electrodes. The current, picked up by electrodes, is amplified in a transimpedance amplifier (TIA), followed by a boxcar averager (BA). Finally, the signal is read by a computer via a data acquisition card (DAQ). {\bf (b)} Sample delay scan of the signal. Green dots are raw data, black line is the data after applying a band-pass filter. See text for details.}
	\label{fig:setup}
\end{figure}

The full characterization of coherent light fields is essential for both the understanding of light-matter interaction and its 
manipulation on time scales down to the period of a field oscillation. Many existing light-field characterization 
methods rely on the determination of the spectral amplitude and phase of the electric field~\cite{Trebino2000}. While the measurement 
of the spectral amplitude is mostly trivial, the determination of the spectral phase is more challenging. This is the reason why most 
existing methods only provide partial phase information, typically leaving the carrier-envelope phase (CEP) 
undetermined~\cite{Kane1993, Iaconis1999, Miranda2012}, thus preventing a complete characterization of the electric field.

An alternative approach to this problem is to measure the electric field in the time domain, as is done with an oscilloscope for electric 
fields up to GHz frequencies. While field-sampling technique based on Auston switches~\cite{Auston1984} or electro optic sampling (EOS)~\cite{Wu1996} have allowed time-domain measurement of THz waves, transposing these techniques into the optical frequency range has long 
remained a serious challenge. The latter was finally met two decades ago, owing to the development of laser sources for near single-cycle 
pulses~\cite{Goulielmakis2004, Mairesse2005} and the introduction of the attosecond streaking technique~\cite{Itatani2002, Kienberger2004}. 

Attosecond streaking relies on the generation of gas-phase high-harmonics (HHG) of the infrared (IR) waveform to be sampled, referred 
to as the streaking pulse in the following. For a near singe-cycle IR pulse with the right carrier-envelope phase, HHG gives rise to a single 
attosecond burst of XUV radiation. This XUV-pulse that acts as a gate, is much shorter than the period of the 
fundamental. After introducing a variable delay between the gating and streaking pulses, both pulses are focused onto a gas target. 

Photoelectrons generated upon XUV ionization of the atoms by the gating pulse are accelerated by the streaking field. This results in 
the so called ponderomotive streaking, i.e. a shift $\mathbf{\Delta p_{e}}$ of the final electron momentum $\mathbf{p_{e}}$ that 
is proportional to the vector potential $\mathbf{A}(t_i)$ of the streaking pulse at the instant $t_i$ of ionization~\cite{Thumm2015}. 
The temporal evolution of the IR-field is inferred from the delay dependent electron momentum usually measured with an electron 
time-of-flight-spectrometer. While it is a powerful technique to time resolve XUV-matter interactions, the original attosecond 
streaking requires a complex high-vacuum setup and its applicability is limited to near single cycle streaking fields.

Several techniques enabling a full characterization of laser pulses have been developed since then. Like attosecond streaking, the 
majority of these field-sampling techniques rely on the use of a short auxiliary pulse providing a sub-cycle nonlinear 
gating~\cite{Kim2013a, Park2018, Saito2018, Keiber2016, Sederberg2020, Hammond2018}. In the TIPTOE approach \cite{Park2018}, 
a strong few-cycle phase-stable gating pulse tunnel ionizes a target gas, while the ionization rate is modulated by the much weaker field 
to be sampled. This modulation can be measured electronically \cite{Park2018} or optically \cite{Saito2018}. More recently, the streaking 
technique was adapted to sample infrared laser fields with a few-cycle visible pulse via ionization of a dilute gas target~\cite{Kubel2017} or 
via electron-hole pair excitation in solid crystals~\cite{Sederberg2020}.

Here, we introduce a novel technique facilitating streaking in ambient air, i.e. the direct measurement of the light field under ambient conditions. Our approach relies on the measurement of transient currents in ambient air plasma, recently 
demonstrated by Kubullek et al.~\cite{Kubullek2020}, and avoids the intermediate HHG step that is essential in attosecond streaking. 
Leaving out the complex, sensitive, and rather inefficient HHG step leads to an order-of-magnitude simpler implementation of the 
streaking concept, while considerably widening its range of applications. In particular, we anticipate that our technique will become an attractive 
tool for the characterization of infrared and mid-infrared pulses generated in novel state-of-the-art laser systems.

\begin{figure*}[ht!]
	\centering\includegraphics{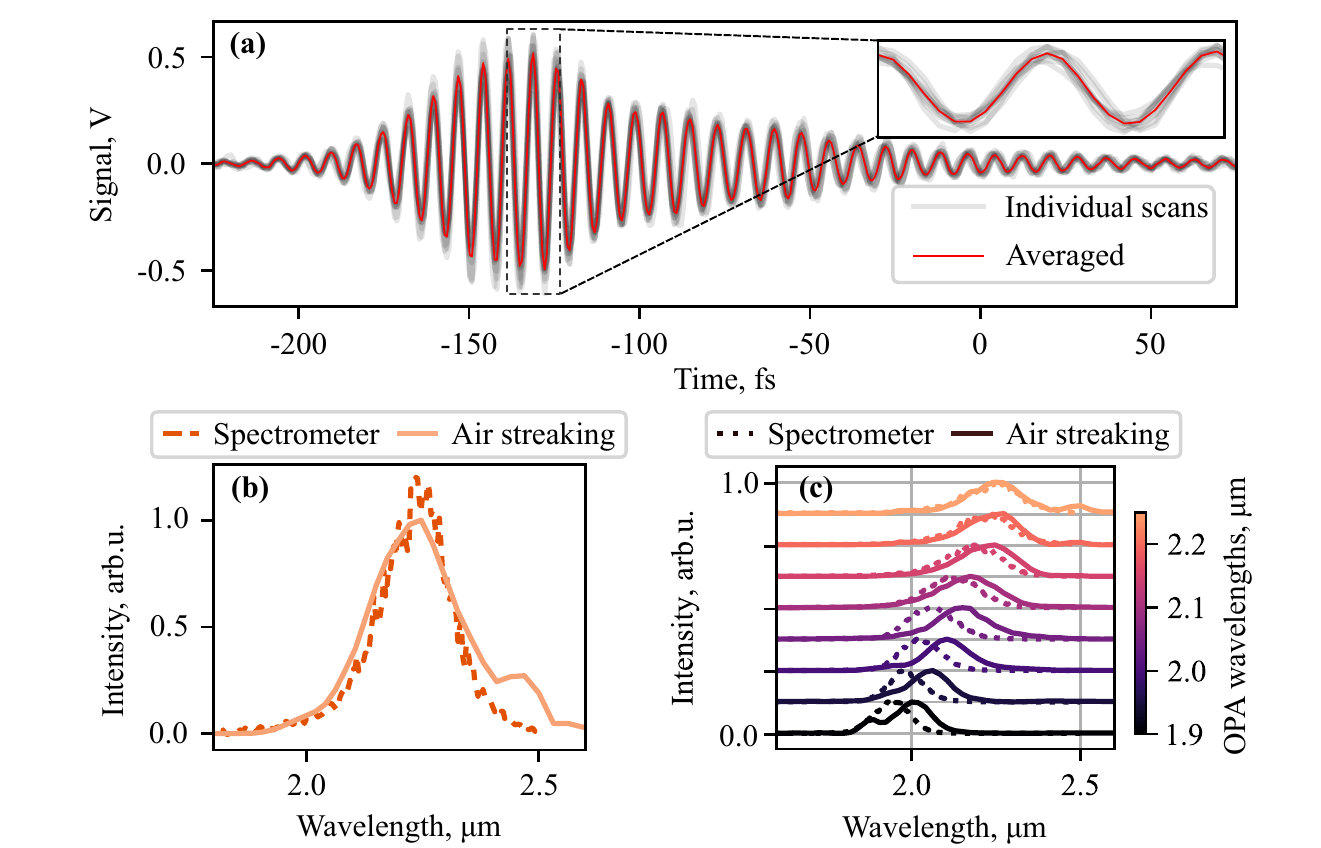}
	\caption{{\bf (a)} Individual filtered air streaking scans (grey), and their average (red). Magnified region of just two laser periods is shown in the inset. {\bf (b)} Spectral intensity of the averaged air streaking trace (solid line) and the spectrum measured with an IR spectrometer (dashed line). {\bf (c)} Air streaking (solid lines) and spectrometer (dashed lines) measurements of OPA pulses with different wavelengths, illustrated by the color code.}
	\label{fig:wldep}
\end{figure*}

\section{Experimental setup}
In our proof of principle experiment, we use a few-cycle 800~nm pulse with non-stabilized CEP as a gating pulse 
to sample a CEP-stable streaking field with a central wavelength at around 2.2~$\upmu$m. 

Both the streaking field and the gating pulse are generated from Coherent Legend Cryo-HE 10 kHz Ti:Sa laser system delivering 1.3 mJ pulses with a duration of 25 fs full-width at half maximum (FWHM) and a central wavelength of 800 nm.

The experimental setup is presented in \figref{fig:setup}{a}. Around 90\% of the output power is used to pump Light Conversion TOPAS-Prime optical parametric amplifier (OPA) producing phase-stable idler streaking pulses with a central wavelength tunable in a spectral range 
from 1.8 to 2.4 $\upmu$m. In order to generate the gating pulses, 10\% of the beam power is split off with a dielectric beam splitter, 
and spectrally broadened inside an Ar-filled hollow-core fiber (HCF). The resulting pulses with a central wavelength at 770 nm are sent 
through a chirped mirror compressor, a 3~mm-thick KDP crystal, and a pair of fused silica wedges, where they are recompressed to a 
FWHM duration of 5 fs. 

The vertically polarized gating pulse (Ti:Sa) and the horizontally polarized streaking pulse (OPA) are attenuated using neutral density filters to $\sim$20~$\upmu$J and $\sim$30~$\upmu$J, respectively. The two beams are then combined on a silicon beam splitter 
(BS) positioned at Brewster's angle for the streaking pulse in order to minimize the reflection losses. A 90$^\circ$ off-axis parabolic silver 
mirror with an effective focal length of 150~mm is used to focus the collinear beams into ambient air and generate a narrow plasma channel. 
The transient polarization induced in the air plasma is probed by measuring the current flowing through a pair of plane parallel 
electrodes placed on either side of the plasma channel. The electrodes, made of 0.5~mm-thick aluminum plates, are separated by a ~0.2~mm 
gap and oriented perpendicular to the polarization of the streaking pulse. The current is amplified in a transimpedance 
amplifier (TIA), and integrated over a time window of 30~$\upmu$s using a boxcar averager (BA) voltage amplifier. The boxcar DC voltage output, 
which is proportional to the charges flowing in the circuit, is recorded with a computer via a data acquisition card (DAQ). In the streaking 
beam path, a delay stage driven with a precision piezo positioner permits controlling the delay between gating and 
streaking pulses with sub-femtosecond precision. 

Two different modes, a fast and a slow mode, can be employed to scan the delay. In the fast mode, the stage is moving at a constant speed 
of 0.5 mm/s throughout the desired scanning range, while the signal is acquired continuously with the DAQ. This mode allowed to scan 
a picosecond range within a few hundreds of milliseconds, facilitating real time observation of the measured waveforms on an oscilloscope at 2 Hz refresh rate. 
Although useful for a quick estimate of the waveform during alignment, we found that measurements using this fast 
mode were flawed due to an uneven motion of the piezo stage at higher speed. For the sake of accuracy, all data presented in the following 
were acquired in the slow mode, where the delay stage was moved in discrete steps of 100 nm (0.67 fs) and the DAQ input was averaged 
over 200 laser shots after each step, stretching a single scan to around 2 minutes. The resulting delay-dependent signal is plotted in \figref{fig:setup}{b}. In order to clean the 
raw signal (green dots) from parasitic noise, it is band-pass filtered within a wavelength range from 0.5 to 4~$\upmu$m, which results in the solid 
black line.

An alternative characterization of the streaking field is performed with a Second Harmonic Generation Frequency-Resolved Optical Gating 
(SHG FROG)~\cite{Kane1993} and a Streak Camera for Strong-Field Ionization (STIER)~\cite{Kubel2017}. For the STIER measurement, half of 
the combined beam is picked off by a D-shaped pick-off mirror and redirected into a standard cold target recoil-ion momentum spectroscopy 
(COLTRIMS) apparatus, while the other half of the beam continues to the air-streaking setup.

A detailed description of the STIER technique can be found in~\cite{Kubel2017}. Briefly, inside the COLTRIMS vacuum chamber, the beams are 
focused onto a supersonic D$_2$ molecular gas jet with a parabolic mirror. The electrons and ions generated in the interaction region are 
guided by homogeneous and parallel electric and magnetic fields onto two time- and position-sensitive detectors. The three-dimensional 
ion momentum reconstructed from the time and position information is measured as a function of the delay between the pulses. 
As demonstrated in~\cite{Kubel2017}, the delay dependent D$_2^+$ momentum is proportional to the vector potential of the streaking field.

\section{Results and discussion}
The measured streaking signal is shown in \figref{fig:wldep}{a}, where the filtered waveform obtained from single scans
(grey lines) is plotted together with an average over 10 scans (red line). In order to correct for slight variations in the 
zero position of the delay stage, the waveforms of the individual scans were shifted along the delay axis as to maximize their 
overlap prior to averaging. The agreement between the single and the averaged waveforms demonstrates the reproducibility of the measurement 
up to the small random delay offsets of order of 1.5~fs. The $\sim$7 fs period of the observed oscillation is in agreement with the 2.2 $\upmu$m 
central wavelength of the streaking field.

\begin{figure}[t]
	\centering\includegraphics{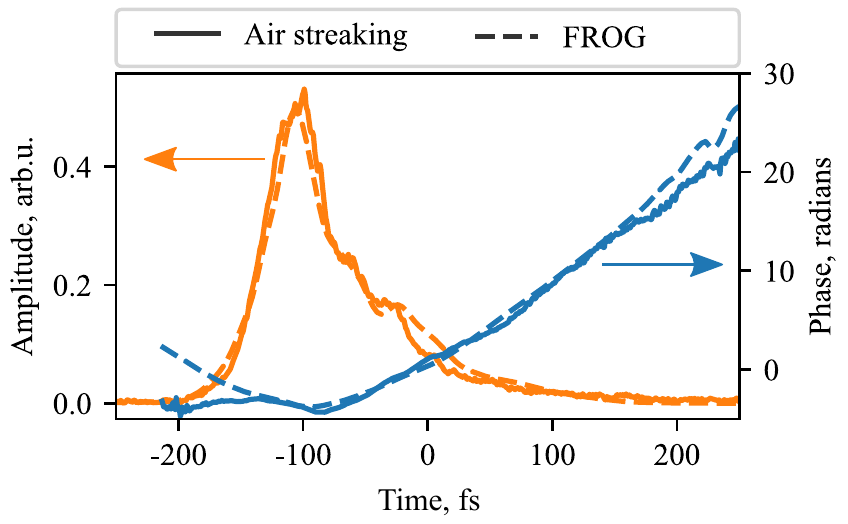}
	\caption{Comparison between the temporal amplitudes (orange) and phases (blue), retrieved with air streaking (solid lines) and SHG FROG (dashed lines).}
	\label{fig:frog}
\end{figure}

In the following, we will show how the measured signal relates to the electromagnetic field of the streaking pulse. To this end, 
we first compare in \figref{fig:wldep}{b} the power spectrum calculated from the measured traces (solid line) with that of the 
streaking pulse measured with an IR spectrometer (dashed line). In \figref{fig:wldep}{c} we show the result of the same comparison for 
different OPA output pulses with central wavelengths ranging from 1.9 to 2.25 $\upmu$m.
The wavelength dependence of the air-streaking and spectrometer measurements follows the same trend. We attribute the discrepancies observed at shorter wavelengths to a jitter in the 
pulse delay that is inherent to our setup, and predominantly affects higher frequencies.

In general, even small noise in the time domain can result in strong variation of the spectral characteristics. In \figrefs{fig:frog}, 
the temporal amplitude and phase of the streaking pulse reconstructed from an independent SHG FROG measurement (dashed red and blue lines, 
respectively), are compared to the amplitude and phase extracted from the air-streaking measurement (solid red and blue lines,
respectively). The CEP left undetermined in the FROG measurement was adjusted to maximize the agreement between both curves. 
As can be deduced from the similarity of the temporal amplitude and phase obtained with both methods, ambient-air streaking and FROG 
lead to comparable results.

\begin{figure}[t!]
	\centering\includegraphics{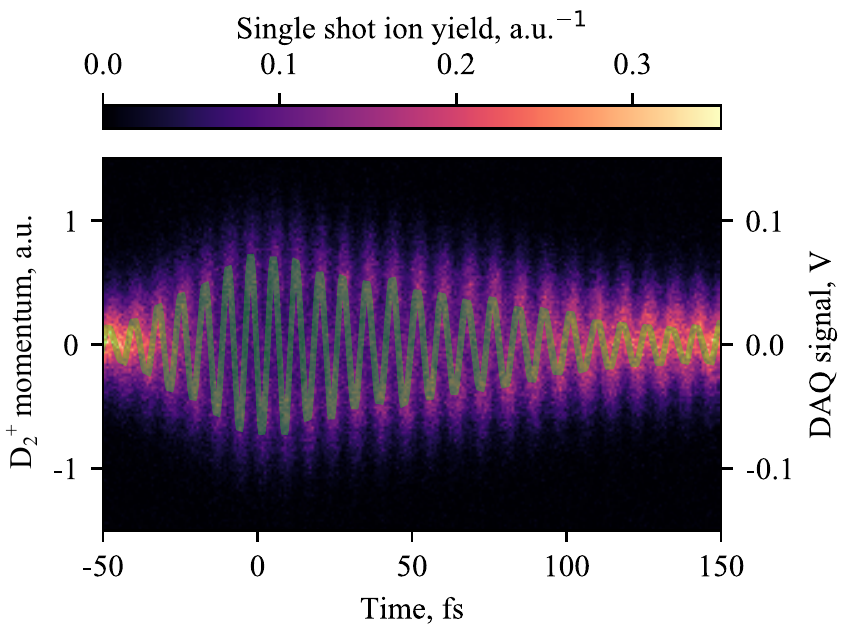}
	\caption{Distribution of the z-component of the D$_2^+$ recoil momentum, measured with the STIER technique (colormap) and the signal acquired  via ambient air streaking measurement and numerically propagated to compensate for the phase difference (green line). The line thickness represents the experimental error.}
	\label{fig:coltrims}
\end{figure}

Despite the good agreement between the two measurements, the ambiguity in the CEP inherent to the FROG technique prohibits a definite statement about the exact nature of the air streaking signal. In regard of the complex collective electron dynamics in the plasma, it is a priori not clear how the phase of the measured current relates to that of the driving field. As argued in~\cite{Kress2006}, electron scattering with neighbouring atoms and ions in ambient air plasma may significantly impact the induced transient current, and one may thus expect a possible effect on its phase. Other studies, in contrast, suggest that the induced current is in phase with the vector potential of the field~\cite{Kim2007}. In order to shed more light onto
the nature of the measured air streaking signal, we performed a parallel measurement of the waveform with the STIER method~\cite{Kubel2017}. The comparison of the two experiments is especially interesting, as STIER is a streaking measurement on a single molecule, and as such provides a clean reference. To account for the slight differences in dispersion along the path from the pick-off mirror to the respective interaction regions in both measurements, the measured air streaking trace was propagated numerically through the extra air in the STIER arm. Additional mirror reflections were also taken into account.

The recorded delay dependent distribution of the ion momentum along the IR polarization axis (in $z$-direction) is shown 
in \figrefs{fig:coltrims} together with the numerically propagated air streaking signal. We find that both signals oscillate essentially in phase, with the ion signal slightly lagging behind by $\Delta\varphi=(500\pm700)~\mathrm{mrad}$. The error mostly results from the uncertainty in propagation lengths in air and optical windows. This finding suggests that ambient-air streaking samples the vector potential. Further studies are needed, however, to confirm and fully elucidate the mechanism that governs the transient current generation.

It is instructive to compare the present method to the related TIPTOE technique~\cite{Park2018}. Similar to air streaking, it probes a current flowing between metal electrodes in ambient air to sample the light waveform. Despite these similarities, there are, 
however, important differences between the two.

One of them concerns the current detection. As shown in Ref.~\cite{Park2018}, provided the bias DC field applied in TIPTOE 
to separate electrons and ions is strong enough, the electrons cross the large air gap and eventually make their way to the electrodes. 
On the other hand, in the recently reported single-shot CEP measurements in ambient air \cite{Kubullek2020} or optoelectronic 
field measurement in solids~\cite{Sederberg2020}, the separated charges recombine before reaching the electrodes, so that the 
measured current is a displacement current.

In ambient air streaking, where no external field is applied, the electron motion is essentially determined by the vector potential 
of the streaking field. As such, at a typical electron momentum transfer cross-section of $10^{-15}~\mathrm{cm^2}$~\cite{Engelhardt1964} 
in air, the induced current is expected to decay after only 
$\lambda\sim1/(2.5\times10^{19}~\mathrm{cm}^{-3}\cdot10^{-15}~\mathrm{cm^2})=400~\mathrm{nm}$ of propagation, i.e. a few orders of magnitude 
less than the electrode separation. This suggests that the current measured in ambient-air streaking is also a displacement current.

In TIPTOE the tunnel ionization yield from the gating pulse is modulated by the electric field to be sampled. Since the sign of the gating 
field changes every half cycle, it is important to concentrate the plasma production to essentially a single half-cycle. As a consequence, 
the gating pulse must be CEP-stabilized with a duration close to a single cycle. Even then, a separate measurement is required to determine 
the absolute direction of the electric field. This is in contrast to the present method, which does not require 
a CEP-stable gating pulse and yet allows for the determination of the CEP of the streaking pulse.

\section{Conclusion}

In conclusion, we have introduced air streaking as a novel technique to measure the electromagnetic field of a coherent light wave. While similar to 
attosecond streaking, our technique avoids the HHG step and works in ambient air,  providing a direct measurement of the vector potential. 
The measured field can be viewed in real time on any standard oscilloscope in the same way as a radio frequency electronic signal. 
The potential of the technique reaches far beyond waveform characterization applications. In a similar way attosecond and solid-state 
streaking have permitted time-resolution of XUV-matter interactions and carrier dynamics in crystals, respectively, ambient air streaking 
opens the door for the study of time-domain sub-femtosecond dynamics inside plasmas.

\section*{Funding}
Air Force Office of Scientific Research (AFOSR) (FA9550-16-1-0109); Canada Research Chairs (CRC); Natural Sciences and Engineering Research Council of Canada (NSERC); National Research Council, Joint Center for Extreme Photonics; German Research Foundation (DFG) (KL-1439/11-1, LMUexcellent); Max Planck Society; European Research Council (ERC) (FETopen PetaCOM).

\section*{Acknowledgments}
We thank David Crane and Ryan Kroeker for their technical support, and are grateful for fruitful discussions  
with Marco Taucer, Guilmot Ernotte, Dmitry Zimin, Nicholas Karpowicz, and Shawn Sederberg.



\bibliography{../../library}

\end{document}